\documentclass[a4paper,twocolumn,fleqn]{article} 
\usepackage{amsmath}                 
\usepackage{txfonts}                   
\usepackage[dvipdfmx]{graphicx}
\usepackage{graphicx,epsfig}
\usepackage{xcolor}
\usepackage{tabularx}
\usepackage{pfr_APiP2025}                 

\begin{document}

\title{Simulation study of neutral tungsten emissions for fusion applications}

\author{Ritu~Dey \sup{1,\#}, Ayushi Agrawal \sup{2,\#}, Reetesh Kumar Gangwar \sup{1}, Deepti Sharma \sup{3}, Rajesh Srivastava \sup{2}, Malay B. Chowdhuri \sup{3} and Joydeep Ghosh \sup{3,4}}

\affiliation{\sup{1}Department of Physics, Indian Institute of Technology Tirupati, Yerpedu 517619, India \\
  \sup{2}Department of Physics, Indian Institute of Technology Roorkee, Roorkee 247667, India \\
  \sup{3}Institute for Plasma Research, Gandhinagar 382428, India\\
  \sup{4}Homi Bhabha National Institute, Training School Complex, Anushakti Nagar, Mumbai 400094, India}

\date{}

\email{ritu.dey@iittp.ac.in}

\begin{abstract}
The article reports electron-impact excitation cross-sections and rate coefficients for neutral tungsten for three transitions (400.87 nm, 429.46 nm, and 430.21 nm) using the relativistic distorted wave approach within the flexible atomic code. Some of these lines are also observed in tokamak plasma. Cross-sections are computed for incident electron energy up to 30 keV. The energy levels in flexible atomic code were corrected to match the NIST database. The electron impact excitation rate coefficients are also provided.\\ \ \\
{\# Contributed equally}

\end{abstract}


\AbstNum{01}


\maketitle

\begin{normalsize}

\newpage
\section{Introduction}
Tungsten is a highly valuable material for magnetic confinement devices due to its low tritium retention, low erosion rate, and high melting point. These properties make it an ideal choice for critical components in fusion reactors. As a result, tungsten is used in the construction of the ITER first wall and divertor plates [1] to handle the high heat fluxes. Similarly, the upgrade version of EAST is used a ITER-like tungsten monoblock divertor configuration [2,3]. Moreover, WEST is a superconducting, actively cooled, full tungsten (W) tokamak [4]. Previously, there were many studies involving erosion due to tungsten [5]. For example, in the divertor of the ASDEX-Upgrade tokamak, erosion properties are investigated by observing the WI emission line at 400.9 nm [5]. It is found that the sputtering yield is in the order of 10$^{-4}$ atoms/ion for
typical divertor plasma conditions. JET was used tungsten on the divertors [6] to handle the
wall materials issues that are critical for the fusion devices. The usage of tungsten material makes it abundant in fusion devices, and it presents as an impurity material. Due to a large radial variation of electron temperature in tokamak devices, various charged states of tungsten along with the neutral can exist inside the tokamak. Over the years, spectroscopic investigations/diagnostics have been carried out to study the W emissions [7] to determine the erosion rate. Various theoretical approaches are used to calculate the energy levels, radiative transition probabilities, and ionization/excitation cross sections, which support the spectroscopic diagnostics results. The various approaches applied worldwide for the above purpose are relativistic distorted wave approximation (RDW) [8], relativistic Hartree-Fock method [9], multi-configuration Dirac-Fock method (MCDF) [10], binary encounter Bethe Model [11], etc. The well-known codes in which these methods are implemented to calculate various atomic structures are Cowan [12,13], flexible atomic code (FAC) [14], DARC [15], GRASP [16], MDFGME [17]. Very recently, Duck-Hee Kwon and Paul Indelicato [18] reported MCDF calculations to obtain radiative electron-impact transition rates for neutral tungsten. They have reported electron impact excitation rate-coefficients for visible transition wavelengths of neutral W of 400.87 nm, 488.69 nm, 498.26 nm, and 522.47 nm for incident electron energy up to 30 eV. They found that the accuracies for electron impact excitation (EIE) cross-sections are challenging, and more experimental and theoretical investigations are necessary to evaluate the neutral W emissions. To address this issue, in the present manuscript, the electron impact excitation cross sections are calculated by using the relativistic distorted wave approximation within the FAC [14] code for the neutral W atom. It is to be noted that these WI emission lines are observed in laser induced breakdown spectrsocopy (LIBS) experiments [19]. To the best of our knowledge, the results reported here are presented for the first time, except for those at the 400.87 nm wavelength. Furthermore, the cross-sections are averaged over electron energy space to obtain the Maxwellian-averaged EIE rate coefficients.
The paper is organized as follows: Section 2 describes the brief theory for calculating the cross-section and rate coefficients within the RDW approximation. Section 3 contains the results. A concluding remark is given in section 4.
\section{Theory}
Flexible atomic code (FAC) [14] Version 1.1.5  is used to simulate the EIE cross-sections. The radial wavefunctions for single-electron orbitals are obtained with a self-consistent field method based on the Dirac equations. The Dirac–Coulomb Hamiltonian is diagonalized to obtain energy levels and atomic state wavefunctions. This code is used the RDW approximation to determine the collision strength for EIE process for neutral W. The collision strength ($\Omega_{01}$) is expressed as,
\begin{eqnarray}
\Omega_{01}=2 \sum_{k} \sum_{{\alpha_{0} \alpha_{1} \beta_{0} \beta_{1}}} Q^{k}(\alpha_0 \alpha_1;\beta_0 \beta_1) \\ \nonumber \langle \Psi_0 ||Z^k(\alpha_0, \alpha_1)||\Psi_1 \rangle \langle \Psi_0 ||Z^k(\beta_0, \beta_1)||\Psi_1 \rangle,
\end{eqnarray}
where,

\begin{eqnarray}
 Q^{k}(\alpha_0 \alpha_1;\beta_0 \beta_1)=\sum_{\kappa_0, \kappa_1}[k]^{-1} P^{k}(\kappa_0 \kappa_1; \alpha_{0} \alpha_{1}) \\ \nonumber P^{k}(\kappa_0 \kappa_1; \beta_{0} \beta_{1}).
\end{eqnarray}
$\kappa_0$, $\kappa_1$ are the relativistic angular quantum numbers of the incident and the
scattered electrons, respectively. $\Psi_0$ and $\Psi_1$ are the wave-functions of the initial and final state, respectively. $Z^k$ ($\alpha_0$ , $\alpha_1$ ), Z$^k$ ($\beta_0$, $\beta_1$) are the tensor operators. $\alpha_0$ , $\alpha_1$, $\beta_0$ and $\beta_1$ are the orbitals of the electron. The radial integral $Q^k$ $(\alpha_0 \alpha_1 ; \beta_0 \beta_1 )$ contains (P$^k$) which is expressed as:
\begin{align}
P^{k}(\kappa_0 \kappa_1; \alpha_0 \alpha_1) &= X^k(\alpha_0 \kappa_0; \alpha_1 \kappa_1) + \sum_t (-1)^{k+t} [k] \nonumber \\
&\quad \times
\begin{Bmatrix}
J_{\alpha_0} & J_1 & t \\
j_0 & j_{\alpha_1} & k
\end{Bmatrix}
 X^t(\alpha_0 \kappa_0;\kappa_1 \alpha_1).
\end{align}
\noindent
$X^k$ is the two electron Slater integral and $k$ is the multipolarity. The total EIE cross-section is expressed in terms of collision strength as (in atomic unit) follows:
\begin{eqnarray}
\sigma_{01}=\frac{\pi}{k_0^2 g_0} \Omega_{01}.
 \end{eqnarray}
Because the RDW theory is perturbative, its accuracy near threshold energies is limited. Therefore, low-energy corrections up to approximately two to three times the excitation threshold are required. The corrected cross section, expressed as a function of the projectile electron energy, is given below [20].
 \begin{eqnarray}
\sigma_{01[corrected]}=[1-(E_{threshold}/E)^3]\sigma_{01},
 \end{eqnarray}
where E is the electron energy and E$_{threshold}$ is the threshold energy for the excitation. The similar equation is applied to find out the electron impact excitation cross-sections and rate coefficients using fully relativistic distorted wave approach of
cesium atom [21].

The electron-impact excitation cross-sections is averaged over the Maxwellian electron energy space and then the rate coefficient is expressed as:
\noindent
\begin{equation}
R_{ij}=2\sqrt{\left(\frac{1}{\pi m_e}\right)}(K_B T_e)^{-3/2}\int_{{\epsilon}}^{\infty}E \sigma_{ij}(E)exp\left(-\frac{E}{K_B T_e}\right) dE,
\end{equation}
where, $K_B$ is the Boltzmann coefficient and m$_e$, T$_e$ are the mass and temperature of the electron, respectively. $\epsilon$ is the threshold energy for the particular transition.

\section{Results and Discussion}
In the present work, FAC is parallelized and executed using 32 CPU cores on ANTYA, an IPR LINUX cluster, to calculate the EIE cross-sections and rate coefficients. It is worth mentioning that the cross-sections and rate coefficients calculated for the particular transitions in the visible range, which are often measured in plasma diagnostics [4, 6], are found in various tokamaks. It is also observed by the LIBS experiment [19]. Neutral tungsten atom which has a total of 74 electrons and the ground state configuration is $[Xe]4f^{14}$ $5d^46s^2$ ($^5D_0$). The atomic structure of neutral tungsten (W I) contains partially filled 5d subshells and due to this the complexity arises in theoretical modelling to investigate the fine-structure spectrum.  Table 1 presents the configurations and excitation energies of the neutral W in lower and upper states from the NIST database [22]. A total of 1754 fine-structure-resolved energy levels are considered, corresponding to the excited level configurations. The energy levels from FAC are corrected to match with the energy values from NIST. Figure 1 displays the collison strength for the above three transitions listed in Table 1. The configurations that are taken for the present calculations are as follows: 5d$^4$6s$^2$, 5d$^5$6[s,p], 5d$^4$6s6[p,d], 5d$^4$6s7[s,p] and 5d$^6$.  It is noticed that all three transitions are dipole allowed transitions and the transition probabilities are 1.63 $\times$10$^7$ s$^{-1}$, 1.24 $\times$10$^7$ s$^{-1}$, and 3.60 $\times$10$^6$ s$^{-1}$, for wavelengths 400.87 nm, 429.46 nm, and 430.21 nm, respectively. It is noticed from figure 1, the collision strengths for the transtions from 5d$^5$6s ($^7S_3$) to 5d$^5$6p ($^7P^o_4$ and $^7P^o_2$), do not vary much ($\sim $ two times) up to incident electron energy of 100 eV. While, transition from 5d$^5$6p to 5d$^4$6s6p ($^7D^o_3$), the collision strength falls 6 times up to 30 eV and beyond that energy value the collision strength is almost constant. Furthermore, the EIE cross-sections that are related to collision strength through equation (4) are evaluated for all three transitions and  plotted in Figure 2. They follow the same pattern of decreasing with the incident electron energy. The cross-sections are simulated up to an incident electron energy of 30 keV. Near the threshold energy, the cross sections are corrected after using equation (5). The maximum magnitudes are 2.75 $\times$ 10$^{-17}$ cm$^2$, 3.13 $\times$ 10$^{-18}$ cm$^2$ at incident electron energy $\sim$ 5.0 eV, 4.5 eV for the transition wavelengths, 400.87 nm, 429.46 nm, respectively, and it falls $\sim$ 10$^3$ times at incident electron energy of 30 KeV. While for 430.21 nm wavelength, the excitation cross-section shows a maximum 2.50 $\times$ 10$^{-18}$ cm$^2$ at $\sim$ 4.0 eV and falls $\sim$ 10$^3$ times at incident electron energy of 30 keV. In addtion, Figure 3 depicts the EIE rate coefficients for the transition wavelengths 400.87 nm, 429.46 nm, and 430.21 nm, respectively. The rate coefficients are obtained after averaging over the Maxwellian electron energy space using equation (6). The electron impact excitation rate coefficients for transition wavelengths 400.87 nm, 429.46 nm are not varying much for the whole energy range considered here (0-300 eV) except near the threshold energies. However, for the wavelength 430.21 nm, the electron impact excitation rate-coefficient falls rapidly upto 50 eV and beyond this energy it is almost constant. It is worth to be mentioned that the EIE rate-coefficients of WI reported by Kwon et al [18] for 400-522 nm range wavelengths and for incident electron energy up to 30 eV. The present simulated EIE rate-coefficient for 400.87 nm transition wavelength, are lower than their Dirac-R matrix calculations for the energy range $\sim$ 0 to 30 eV.

\begin{table}
\caption{Configurations and energy levels of WI transitions from the NIST [22] recommended values based on experiment.}

\begin{tabular}{|c|c|c|c|c|} \hline
 \textbf{Wavel.} &  \textbf{Config.} &  \textbf{Config.} & \textbf{E$_i$}  & \textbf{E$_k$}  \\
 & Lower & Upper & & \\
(nm) &level &level & (eV) & (eV) \\ \hline
400.87 & 5d$^5$($^6$S)6s ($^7S_3$)&5d$^5$($^6$S)6p ($^7P^0_4$) &0.365 & 3.457\\ \hline
429.46 & 5d$^5$($^6$S)6s ($^7S_3$)&5d$^5$($^6$S)6p ($^7P^0_2$) & 0.365& 3.252\\ \hline
 430.21 & 5d$^5$($^6$S)6s ($^7S_3$)&5d$^5$6s($^6D$)6p ($^7D^0_3$) & 0.365& 3.247\\ \hline
\end{tabular}
\end{table}

\begin{figure}[ht]
  \centering
  \includegraphics[width=7cm]{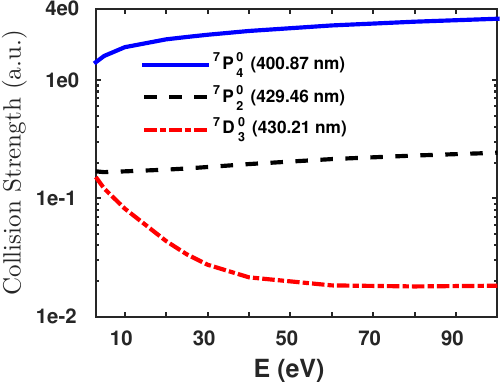}
  \caption{Simulated EIE collison strengths for the excitation from 5d$^5$($^6$S)6s ($^7S_3$)to 5d$^5$($^6$S)6p ($^7P^0_4$) (blue-solid curve), 5d$^5$($^6$S)6p ($^7P^0_2$) (black dashed curve) and 5d$^5$6s($^6D$)6p ($^7D^0_3$) (red dash-dotted curve), respectively.}
  \label{fig:Fig1}
\end{figure}

\begin{figure}[ht]
  \centering
  \includegraphics[width=7cm]{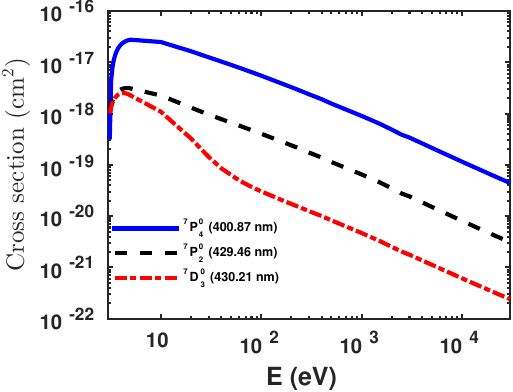}
  \caption{Simulated EIE cross-sections for the excitation from 5d$^5$($^6$S)6s ($^7S_3$)to 5d$^5$($^6$S)6p ($^7P^0_4$) (blue-solid curve), 5d$^5$($^6$S)6p ($^7P^0_2$) (black dashed curve) and 5d$^5$6s($^6D$)6p ($^7D^0_3$) (red dash-dotted curve), respectively.}
  \label{fig:Fig2}
\end{figure}

\begin{figure}[ht]
  \centering
  \includegraphics[width=7cm]{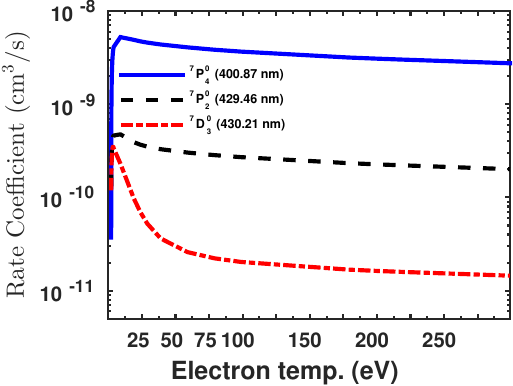}
  \caption{Notations are similar to the figure 1, simulated EIE rate-coefficients for the neutral W.}
  \label{fig:Fig3}
\end{figure}

\begin{figure}[ht]
  \centering
  \includegraphics[width=9cm]{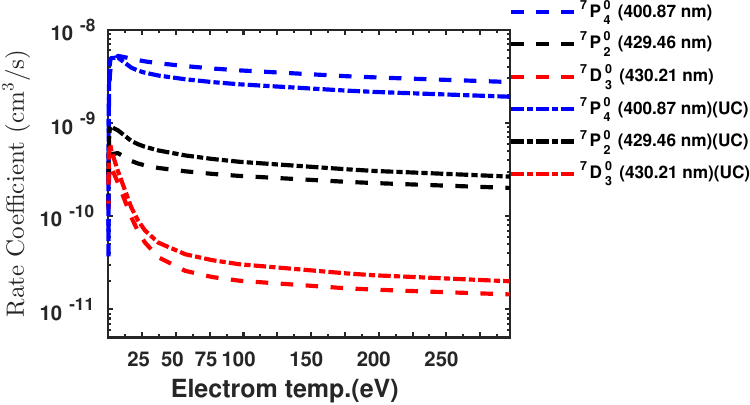}
  \caption{Simulated EIE rate coefficients for the excitation from 5d$^5$($^6$S)6s ($^7S_3$)to 5d$^5$($^6$S)6p ($^7P^0_4$) (blue-dashed curve), 5d$^5$($^6$S)6p ($^7P^0_2$) (black dashed curve) and 5d$^5$6s($^6D$)6p ($^7D^0_3$) (red dashed curve), respectively for corrected energy levels. The blue, black and red dash-dotted curves represent the FAC simulated EIE rate coefficients uncorrected energy levels.                                                                                                                            }
  \label{fig:Fig4}
\end{figure}
\begin{table}
\caption{Energy levels of WI transitions from FAC and the NIST [22] recommended values based on experimental data.}

\begin{tabular}{|c|c|c|c|c|} \hline
 \textbf{Wavel.} &  \textbf{E$_i$} (FAC) &  \textbf{E$_k$} (FAC) & \textbf{E$_i$} (NIST) & \textbf{E$_k$} (NIST)  \\
 & Lower & Upper & Lower & Upper\\
(nm) & (eV) & (eV) & (eV) & (eV) \\ \hline
400.87 & 2.66 & 4.01 &0.365 & 3.457\\ \hline
429.46 & 2.66 & 3.70 & 0.365& 3.252\\ \hline
 430.21 & 2.66 & 2.07 & 0.365& 3.247\\ \hline
\end{tabular}
\end{table}

Finally, Figure 4 illustrates the EIE rate coefficients for the above mentioned neutral W wavelengths, simulated with and without corrected excited energy levels. The energy levels before correction for these three particular transitions are listed in Table 2. It is noticed that, the maximum deviation in EIE rate-coefficient are almost $\sim$ 50$\%$ for all the three transitions ((5d$^5$($^6$S)6s ($^7S_3$)to -5d$^5$($^6$S)6p ($^7P^0_4$),- 5d$^5$($^6$S)6p ($^7P^0_2$) and -5d$^5$6s($^6D$)6p ($^7D^0_3$)) considered here. It is hereby confirmed that energy level corrections are necessary to obtain the accurate collision strengths and cross sections for the neutral W.

\section{Conclusion}
The electron impact excitation cross-sections for WI are simulated with RDW approximation within the FAC code. The Maxwellian averaged rate coefficients are also presented. The wavelengths which are considered presently are, 400.87 nm, 429.46 nm, and 430.21 nm which correspond to transition from 5d$^5$($^6$S)6s ($^7S_3$) to 5d$^5$($^6$S)6p ($^7P^0_4$), 5d$^5$($^6$S)6p ($^7P^0_2$) and 5d$^5$6s($^6D$)6p ($^7D^0_3$), respectively. The electron impact excitation rate coefficients for transitions in the visible range are presented, which will contribute significantly to spectroscopic diagnostics in fusion devices.
\end{normalsize}

\section{Acknowledgements}
The authors, Ritu Dey, Reetesh Kumar Gangwar, Deepti Sharma, Rajesh Srivastava, and Joydeep Ghosh acknowledge the Board of Research in Nuclear Sciences (BRNS), DAE, Government of India, for supporting this work under research project grant (sanction no. 57/14/01/2023/12089). One of the author, Ritu Dey acknowledges Dr. Ankit Dhaka of the Institute for Plasma Research (IPR) for his assistance in running the parallelized FAC code on ANTYA, the IPR Linux Cluster. The simulation results presented in this study were obtained using ANTYA, the IPR Linux Cluster.

\end{document}